%Paper: hep-ph/9405431
%From: manohar@sphal.UCSD.EDU (Aneesh V. Manohar)
%Date: Tue, 31 May 1994 21:01:03 -0700

%%%%%%%%%%%%%%%%%%%%%%%%%%%%%%%%%%%%%%%%%%%%%%%%%%%%%%%%%%%%%%%%%%%
%                       INSTRUCTIONS
%
% This paper uses the harvmac macros. 1 postscript figure
% has been included as a uuencoded tar file with instructions for
% unpacking. If you have  epsf.tex, uncomment the following line
% and the postscript figure
%
%\input epsf
%
% will be included in the paper by the dvips program. If you do not
% have epsf.tex, you can print the figures out separately.
%
%%%%%%%%%%%%%%%%%%%%%%%%%%%%%%%%%%%%%%%%%%%%%%%%%%%%%%%%%%%%%%%%%%%%%%%%
\ifx\epsffile\undefined\message{(FIGURES WILL BE IGNORED)}
\def\insertfig#1{}% null macro
\else\message{(FIGURES WILL BE INCLUDED)}
\def\insertfig#1{{{
\midinsert\centerline{\epsfxsize=\hsize
\epsffile{#1}}\bigskip\bigskip\bigskip\bigskip\endinsert}}}
\fi

\input harvmac

%%%%%%%%%%%%%%%%%%%%%%%%%%%%%%%%%%%%%%%%%%%%%%%%%%%%%%%%%%%%%%%%%%%%%%
%
%  UCSD macros to overwrite some of the definitions in harvmac.tex
%  (include after harvmac.tex)
%  last modified 4/92
%
%%%%%%%%%%%%%%%%%%%%%%%%%%%%%%%%%%%%%%%%%%%%%%%%%%%%%%%%%%%%%%%%%%%%%%%
%
% modify the output routine for the little format
%
\ifx\answ\bigans
\else
\output={
  \almostshipout{\leftline{\vbox{\pagebody\makefootline}}}\advancepageno
}
\fi
%
%
% address
%

%
% grant numbers
%

%
% preprint number
%
\def\UCSD#1#2{\noindent#1\hfill #2%
\bigskip\supereject\global\hsize=\hsbody%
\footline={\hss\tenrm\folio\hss}}% restores pagenumbers
%
% abstract
%
\def\abstract#1{\centerline{\bf Abstract}\nobreak\medskip\nobreak\par #1}
%
%
% titlefont
%
%
\edef\tfontsize{ scaled\magstep3}
 \tfontsize  \tfontsize
 \tfontsize \font\titlei=cmmi10 \tfontsize
\font\titleis=cmmi7 \tfontsize \font\titleiss=cmmi5 \tfontsize
\font\titlesy=cmsy10 \tfontsize \font\titlesys=cmsy7 \tfontsize
\font\titlesyss=cmsy5 \tfontsize  \tfontsize
\skewchar\titlei='177 \skewchar\titleis='177 \skewchar\titleiss='177
\skewchar\titlesy='60 \skewchar\titlesys='60 \skewchar\titlesyss='60
%
%\def\titlefont{\def\rm{\fam0\titlerm}% switch to title font
%\textfont0=\titlerm \scriptfont0=\titlerms \scriptscriptfont0=\titlermss
%\textfont1=\titlei \scriptfont1=\titleis \scriptscriptfont1=\titleiss
%\textfont2=\titlesy \scriptfont2=\titlesys \scriptscriptfont2=\titlesyss
%\textfont\itfam=\titleit \def\it{\fam\itfam\titleit}\rm}
%
%
% math symbols
%
%---------------------------------------------------------------------
%
\def\inv{^{\raise.15ex\hbox{${\scriptscriptstyle -}$}\kern-.05em 1}}
  %prime
\def\lbar{{\lower.35ex\hbox{$\mathchar'26$}\mkern-10mu\lambda}} %lambda bar

%
%
% various slashed symbols
%
%
 % slashes a character
\def\dsl{\,\raise.15ex\hbox{/}\mkern-13.5mu D} %this one can be subscripted
\def\delsl{\raise.15ex\hbox{/}\kern-.57em\partial}
\def\Ksl{\hbox{/\kern-.6000em\rm K}}
\def\Asl{\hbox{/\kern-.6500em \rm A}}
\def\Dsl{\hbox{/\kern-.6000em\rm D}} %roman D
\def\Qsl{\hbox{/\kern-.6000em\rm Q}}
\def\gradsl{\hbox{/\kern-.6500em$\nabla$}}
%
% space and backspace in l mode
%
\def\lspace{\ifx\answ\bigans{}\else\qquad\fi}
\def\lbspace{\ifx\answ\bigans{}\else\hskip-.2in\fi} % $$\lbspace...$$
%
%     boxes an equation
%
\def\boxeqn#1{\vcenter{\vbox{\hrule\hbox{\vrule\kern3pt\vbox{\kern3pt
        \hbox{${\displaystyle #1}$}\kern3pt}\kern3pt\vrule}\hrule}}}
%
%     draw a little box (end of proof symbol)
%     e.g. \mbox{.1}{.1}
%
\def\mbox#1#2{\vcenter{\hrule \hbox{\vrule height#2in
\kern#1in \vrule} \hrule}}
%
%
%
%     curly letters
%
   %curly letters

\def\CM{{\cal M}} \def\CN{{\cal N}} \def\CO{{\cal O}}

%
%
%
%     derivatives
%
%

%

\def\bar#1{\overline{#1}}

\def\bra#1{\left\langle #1\right|}
\def\ket#1{\left| #1\right\rangle}

\def\darr#1{\raise1.5ex\hbox{$\leftrightarrow$}\mkern-16.5mu #1}

%
 %pound sterling
%
 %puts a small half in a displayed eqn
\def\frac#1#2{{\textstyle{#1\over #2}}} %puts a small fraction
%in a displayed eqn
%
%
%     various math operators
%
%

%
%
%
%

%
%       relations
%
\def\ltap{\ \raise.3ex\hbox{$<$\kern-.75em\lower1ex\hbox{$\sim$}}\ }
\def\gtap{\ \raise.3ex\hbox{$>$\kern-.75em\lower1ex\hbox{$\sim$}}\ }
\def\gl{\ \raise.5ex\hbox{$>$}\kern-.8em\lower.5ex\hbox{$<$}\ }
\def\roughly#1{\raise.3ex\hbox{$#1$\kern-.75em\lower1ex\hbox{$\sim$}}}
%
%
%       This defines et al., i.e., e.g., cf., etc.

%

%
\def\np#1#2#3{Nucl. Phys. B{#1} (#2) #3}
\def\pl#1#2#3{Phys. Lett. {#1}B (#2) #3}
\def\prl#1#2#3{Phys. Rev. Lett. {#1} (#2) #3}
\def\physrev#1#2#3{Phys. Rev. {#1} (#2) #3}

\relax

\noblackbox

\def\clebsch#1#2#3#4#5#6{\left(\matrix{#1&#3\cr#2&#4\cr}\right.\left|
\matrix{#5\cr#6\cr}\right)}

\def\sixj#1#2#3#4#5#6{\left\{\matrix{#1&#2&#3\cr#4&#5&#6\cr}\right\}}

\centerline{{\titlefont{Baryon Magnetic Moments}}}
\medskip
\centerline{{\titlefont{in the $1/N_c$ Expansion}}}
\bigskip
\centerline{Elizabeth Jenkins and Aneesh V.~Manohar}
\smallskip
\centerline{{\sl Department of Physics, University of California at San
Diego, La Jolla, CA 92093}}
\bigskip
\vfill
\abstract{Relations among the baryon magnetic and transition magnetic moments
are derived in the $1/N_c$ expansion.  Relations which hold to {\it all orders}
in $SU(3)$ breaking and to leading and first subleading orders in the $1/N_c$
expansion are obtained.  Additional relations are found which are valid up to
$SU(3)$ breaking at first subleading order in the $1/N_c$ expansion.  The
experimental accuracy of these relations fits the pattern predicted by the
$1/N_c$ expansion.  The predictions of the $1/N_c$ expansion are compared in
detail with those of the non-relativistic quark model.  The $1/N_c$ expansion
explains why certain quark model relations work to greater accuracy than
others.
}
\vfill
%\draftmode
\UCSD{\vbox{\hbox{UCSD/PTH 94-10}\hbox{hep-ph/9405431}}}{May 1994}

The $1/N_c$ expansion of 't~Hooft~\ref\thoofti{G.~'t~Hooft, \np {72} {1974}
{461}, \np {75} {1974} {461}} is one of the few calculational techniques for
obtaining rigorous nonperturbative information about hadrons in QCD.  The
implications of the $1/N_c$ expansion of QCD for baryons were initially worked
out by Witten~\ref\witten{E.~Witten, \np {160} {1979} {57}}.  Recently,
progress has been made in obtaining quantitative results for baryons using the
$1/N_c$ expansion~\ref\dm{R.~Dashen and A.V.~Manohar, \pl {315} {1993} {425,
438}}\ref\ej{E.~Jenkins, \pl {315} {1993} {431, 441, 447}}\ref\djm{R.~Dashen,
E.~Jenkins and A.V.~Manohar, \physrev {D49}{1994}{4713}}.  The baryon sector of
QCD possesses has a contracted spin-flavor symmetry in the large $N_c$
limit~\ref\gs{J.-L.~Gervais and B.~Sakita, \prl{52} {1984} {527}, \physrev {30}
{1984} {1795}.}\dm.  Deviations from exact spin-flavor symmetry can be studied
systematically in the $1/N_c$ expansion by computing $1/N_c$ corrections to the
large $N_c$ limit \dm\ej\djm.  It has been proven that the first non-trivial
correction to ratios of baryon axial vector couplings and of isovector magnetic
moments arises at order $1/N_c^2$ \dm\djm, which accounts for the  success of
the $N_c\rightarrow\infty$ predictions for these quantities at the $10\%$
level.  In ref.~\djm, the implications of the $1/N_c$ expansion for baryons
containing strange quarks are analyzed without assuming $SU(3)$ symmetry.  The
results obtained are therefore valid to all orders in $SU(3)$ breaking, and can
be used to constrain the form of $SU(3)$ breaking for baryons.  The $1/N_c$
expansion justifies a number of results found phenomenologically in baryon
chiral perturbation theory~\ref\chiralrefs{ E.~Jenkins and A.V.~Manohar, \pl
{259} {1991} {353}, \pl{255} {1991} {558}; E.~Jenkins, \np {375} {1992} {561},
\np{368} {1992} {190}}, and provides insight into the successes of
phenomenological models, such as strong coupling theory, the Skyrme model, and
the non-relativistic quark model.

Recent work related to refs.~\dm\ej\djm\ can be found in
refs.~\ref\cgo{C.~Carone, H.~Georgi, and  S.~Osofsky, \pl {322} {1994}
{227}}\ref\lmr{M.~Luty and J.~March-Russell, LBL preprint LBL-34778 {\tt
[hep-ph/9310369]}}\ref\broniowski{ W.~Broniowski, Regensberg preprint
TPR-93-39, {\tt [hep-ph/9402206]}}. Carone, Georgi and Osofsky realized that
spin-flavor symmetry for the low lying baryon multiplets follows from the spin
independence of the baryon wavefunctions, and they extended Witten's
Hartree-Fock analysis to baryons containing light quarks. They also noted that
$SU(6)$ symmetry relations should hold for the isoscalar axial vector currents.
Luty and March-Russell have rederived some of the results in ref.~\dm\ej\djm\
by a different method.  Broniowski has proven that the $1/N_c$ correction to
the isovector magnetic moments vanishes for two flavors using low energy QCD
sum rules.

In this paper, we use the results of refs.~\dm\ej\djm\ to obtain relations
among the baryon magnetic and transition magnetic moments in the $1/N_c$
expansion. Our formul\ae\ for the isovector magnetic moments to leading and
first subleading order in the $1/N_c$ expansion, and for the isoscalar magnetic
moments to leading order in the $1/N_c$ expansion, were derived in
ref.~\dm\djm. The $1/N_c$ corrections to the isoscalar magnetic moments are
determined in this paper.  The predictions we obtain in the $1/N_c$ expansion
are compared in detail with those of the non-relativistic quark model. The
$1/N_c$ expansion gives as many relations among the baryon magnetic moments as
the non-relativistic quark model.  Many of the relations are identical to those
of the quark model, but a few differ.  The $1/N_c$ relations which differ from
the quark model are in better agreement with experiment than the corresponding
quark model relations. In addition, there are some very important advantages to
the $1/N_c$ expansion:  it is an expansion in QCD, which does not make use of
any model description of baryons, and it provides an explanation of why some
relations work better than others. Relations which are true up to corrections
of order $1/N_c$ work to about 30\%, whereas relations which are true up to
order $1/N_c^2$ work to about 10\%.

The $1/N_c$ analysis of this work is complementary to recent calculations of
the baryon magnetic moments in chiral perturbation theory~\ref\jlsm{E.~Jenkins,
M.E.~Luke, M.J.~Savage, and A.V.~Manohar, \pl {302} {1993}
{482}.}\ref\bss{M.N.~Butler, M.J.~Savage, and R.P.~Springer, UCSD-PTH 93-22
{\tt [hep-ph/9308317]}.}\ref\bssii{ M.N.~Butler, M.J.~Savage, and
R.P.~Springer, \pl {304} {1993} {353}}. The $1/N_c$ results are an expansion in
$1/N_c$ valid to all orders in the strange quark mass $m_s$, whereas the chiral
perturbation theory results are an expansion in $m_s$ (including non-analytic
terms) valid to all orders in $1/N_c$.  The relations we derive to second order
in the $1/N_c$ expansion are satisfied to all orders in $SU(3)$ breaking.
Additional relations are derived which are valid to all orders in $SU(3)$
breaking at leading order in the $1/N_c$ expansion, and at first subleading
order in the $1/N_c$ expansion in the $SU(3)$ limit.

In the large $N_c$ limit, baryons are infinitely heavy and can be treated as
static fermions. For static baryons, the axial vector and magnetic moment
operators $\bar B T^a\gamma^\mu\gamma_5 B$ and $\bar BT^a \sigma_{\mu\nu} B$
are both proportional to the spin-flavor operator  $\bar B T^a \vec{\sigma} B$,
where $T^a$ is a flavor matrix.  As a result, the large $N_c$ consistency
conditions for the baryon magnetic moments have the same form as the
consistency conditions for the axial couplings \dm\djm.  The solution of the
consistency conditions for two flavors is given in ref.~\dm\ and the solution
for three flavors is given in ref.~\djm. One subtlety of the three-flavor
analysis is that baryon flavor representations change with $N_c$.  This
complication leads to an ambiguity in identifying baryons for $N_c=3$ with
large $N_c$ baryon states.  The flavor $SU(3)$ weight diagram for spin-1/2
baryons in the $N_c \rightarrow \infty$ limit is given in \fig\weight{The
weight diagram for the spin-1/2 baryons for large $N_c$ QCD. The representation
reduces to the familiar baryon octet for $N_c=3$.}. In ref.~\djm, the
$N_c\rightarrow\infty$ limit is taken with the isospin, spin and strangeness of
the baryon held fixed. Thus, the baryon states of physical interest are
located at a fixed distance from the top of the weight diagram as $N_c
\rightarrow \infty$. With this limiting procedure, one can show that baryon
matrix elements are given in terms of an expansion in $I/N_c$, $J/N_c$,
$J_s/N_c$ and $K/N_c$, where $I$ is the isospin, $J$ is the spin, $J_s$ is the
strange quark spin, and $K= -S/2$, where $S$ is the strangeness. The results of
the $1/N_c$ expansion for arbitrary $SU(3)$ breaking follow from a $SU(4)\times
SU(2)\times U(1)$ spin-flavor symmetry\foot{$SU(4)$ is the spin-flavor symmetry
for two flavors, $SU(2)$ is the strange quark spin symmetry, and $U(1)$ is the
strangeness.} for baryons in the $N_c\rightarrow\infty$ limit.

The baryon magnetic moments are proportional to the quark charge matrix
$$
   Q = \pmatrix{\frac23&\hphantom{-}0&\hphantom{-}0\cr
   \noalign{\smallskip} 0&-\frac13&\hphantom{-}0\cr
   \noalign{\smallskip} 0&\hphantom{-}0&-\frac13\cr}.
$$
$Q$ is a linear combination of the flavor generators $T^3$ and $T^8$, which
transform as an isovector and an isoscalar, respectively. In the large
$N_c$ limit, the isovector magnetic moments are of order $N_c$, and the
isoscalar magnetic moments are of order one (provided one takes the large-$N_c$
limit with $I$, $J$ and $K$ held fixed). The isovector magnetic moments ($\vec
\mu_V$) have the same form as the pion-baryon couplings \dm\djm,
\eqn\isovectori{
\mu^i_V = N_c \mu(K) X_0^{i3}\Biggl[1 + \CO\left( {1 \over N_c^2}
   \right) \Biggr],
}
where $X_0^{ia}$ is the generator of the contracted spin-flavor algebra for
baryons defined in refs.~\dm\djm, and $\mu(K)$ is an unknown coefficient which
is a constant to leading order in $1/N_c$ and is at most linear in $K$ at order
$1/N_c$,
\eqn\muk{
   \mu(K) = \mu_0  + {1\over N_c} \mu_1 K + \CO\left( {1 \over N_c^2} \right),
}
where $\mu_0$ has both order one and $1/N_c$ terms.
The matrix elements of $\mu^i_V$ between baryons labeled by isospin, spin and
strangeness quantum numbers $(I,I_3)$, $(J,J_3)$ and $K$ were worked out
explicitly in ref.~\djm,
\eqn\isovector{\eqalign{
   &\bra{I'\,I_3',J'\,J_3';K}\mu^{i}_V\ket{I\,I_3,J\,J_3;K} = \Biggl[ N_c
   \,\mu(K) \,(-1)^{2J'+J-I'-K}\times\cr
   &\sqrt{(2I+1)(2J+1)}\sixj 1 I {I'} K {J'} J
   \clebsch I {I_3} 1 3 {I'} {I_3'} \clebsch J {J_3} 1 i {J'} {J_3'}\Biggr]\cr
   &\qquad\qquad\qquad\qquad\times\Biggl[1 + \CO\left( {1 \over N_c^2}
   \right) \Biggr].
}}
Similarly, the isoscalar magnetic moments ($\vec \mu_S$) have the same form
as the $\eta$-baryon coupling \djm,
\eqn\isoscalar{
   \mu^{i}_S = a(K) J^i + b(K) J^i_s + \CO\left( {1 \over N_c^2}\right).
}
where $J^i$ is the total baryon spin, and $J_s^i$ is the ``spin of the
strange quarks.''\foot{The operator $J_s$ has a precise meaning in terms of the
induced representations discussed in ref.~\djm. It reduces to the spin of the
strange quarks in a non-relativistic quark model description for the
baryons.} The coefficients $a(K)$ and $b(K)$ are constants to leading order in
$1/N_c$ and are at most linear in $K$ at order $1/N_c$,
\eqn\abexp{\eqalign{
   a(K) &= a_0 + {1\over N_c}\, a_1 K + \CO\left( {1 \over N_c^2} \right),\cr
   b(K) &= b_0 + {1\over N_c}\, b_1 K + \CO\left( {1 \over N_c^2} \right),\cr
}}
where $a_0$ and $b_0$ contain both order one and $1/N_c$ terms.
The order one result for the isoscalar magnetic moments was derived in
ref.~\djm.  Eq.~\abexp\ also contains the new result that the $1/N_c$
corrections to the isoscalar magnetic moments are obtained simply by including
a linear in $K$ term in the coefficients $a(K)$ and $b(K)$ at order $1/N_c$.
No other new operator structures appear at this order. It is important to
emphasize that eqs.~\isovectori--\abexp\ were derived without assuming $SU(3)$
flavor symmetry, and are therefore valid to all orders in $SU(3)$ breaking.
Thus, to leading and first subleading order in the $1/N_c$ expansion, the
isovector magnetic moments are parametrized in terms of two constants $\mu_0$
and $\mu_1$ whereas the isoscalar magnetic moments are parametrized by four
constants $a_0$, $a_1$, $b_0$ and $b_1$.  In the limit of exact $SU(3)$ flavor
symmetry, $\mu_1/\mu_0 = -2$, $a_0/\mu_0 = -4 (\alpha + \beta/N_c)/3\sqrt3$,
$a_1/\mu_0=4\sqrt3+8\alpha/\sqrt3$, $b_0 / \mu_0 = -2\sqrt{3}+4\sqrt3/N_c$, and
$b_1=0$.  These limiting values can be derived using the $SU(3)$ tensor
analysis method of ref.~\djm, and the $1/N_c$ expansion of the $SU(3)$
invariant amplitudes $\CM$ and $\CN$ defined there,
$\CN/\CM=1/2+\alpha/N_c+\beta/N_c^2$.

It is useful to compare the predictions of the $1/N_c$ expansion with those of
the non-relativistic quark model. The quark model predictions for the baryon
magnetic moments are well-known.  The magnetic moment operator for the quarks
is given by
$$
\mu = \mu_u J_u + \mu_d J_d + \mu_s J_s,
$$
where $J_q$ is the spin operator of quark $q$ and $\mu_q$ is its magnetic
moment.  The baryon magnetic moments are obtained by taking the matrix elements
of the magnetic moment operator $\mu$ between baryon states with $SU(6)$
symmetric quark model wavefunctions. $SU(3)$ breaking in the quark model arises
from explicit $SU(3)$ breaking in the quark magnetic moments.  In the isospin
limit, $\mu_u=-2\mu_d$, but $\mu_s$ is unrelated to $\mu_d$.

The matrix elements of $X_0^{i3}$, $J^i$ and $J_s^i$ times the unknown
coefficient functions $\mu(K)$, $a(K)$ and $b(K)$, respectively, describe the
isovector and isoscalar magnetic moments to leading and first subleading orders
in the $1/N_c$ expansion.  The matrix elements of the quark spin operators
$J_q^i$ times the unknown quark magnetic moments $\mu_q$ describe the baryon
magnetic moments in the non-relativistic quark model. The matrix elements of
the above operators are given in Table~1 for all of the octet, decuplet and
decuplet-octet transition magnetic moments.  (In Table~1, the magnetic moment
of baryon $B$ is denoted by $B$ and the $B_1 \rightarrow B_2$ transition
magnetic moment is denoted by $B_2 B_1$.)  The matrix elements listed in
Table~1 are the matrix elements of the $i=3$ (or $\hat z$) components of the
operators between $J_3=1/2$ states for the octet magnetic moments, between
$J_3=3/2$ states for the decuplet magnetic moments, and between $J_3=1/2$
states of both the spin-1/2 and spin-3/2 baryons for the transition magnetic
moments.  The matrix elements of $X_0^{33}$ are given in eq.~\isovector.  The
operator $J^3$ is the total angular momentum, and has only diagonal matrix
elements. The matrix elements of the operator $J_s^3$ can be computed using its
definition in terms of the induced baryon representations given in ref.~\djm.
The matrix elements of the quark model operators are obtained using $SU(6)$
quark model wavefunctions for the octet and decuplet baryons.

There are 27 magnetic moments listed in Table~1. Isospin invariance gives six
linear relations I1--I6, as listed in Table~2, leaving 21 independent magnetic
moments. These 21 magnetic moments can be divided into 11 isovector
combinations such as $(p-n)$ and 10 isoscalar combinations such as $(p+n)$.
Linear relations among the magnetic moments can be derived in the $1/N_c$
expansion.  These relations are listed in Table~3.  The accuracy with which the
relations are satisfied is also tabulated for those relations involving
magnetic moments which have been measured \ref\pdg{K. Hikasa et al., \physrev
{D45} {1992} {1}.}. Since the dominant uncertainty in the magnetic moment
relations is theoretical, not experimental, the relations in Table~3 are
written so that all terms on one side of an equation have the same sign in
order to avoid misleading cancellations.  The accuracy listed in the last
column of Table~3 is the difference between the two sides of each equation,
i.e. $|(lhs-rhs)/(lhs+rhs)/2|$.

The isovector magnetic moments depend on the two unknown parameters $\mu_0$ and
$\mu_1$ in $\mu(K)$ up to corrections of order $1/N_c^2$ relative to the
leading (order $N_c$) term.  Thus, there are nine relations among the 11
isovector magnetic moments (V1--V7, V8${}_1$, V9${}_1$) which are valid to
relative order $1/N_c^2$ and to all orders in $SU(3)$ breaking.  The parameter
$\mu_1$ arises at first subleading order in the $1/N_c$ expansion, so there is
one additional relation among the isovector magnetic moments (V10${}_1$) which
is true at leading order, but which is not satisfied at first subleading order.
 In the $SU(3)$ limit, $\mu_1$ is related to $\mu_0$, so this relation can be
improved by including the $SU(3)$-symmetric $1/N_c$ correction.  There are two
equivalent forms for this improved relation (V10${}_2$ and V10${}_3$), each of
which is satisfied by the leading term to all orders in $SU(3)$ breaking and by
the subleading term in the $SU(3)$ limit.  The two forms, V10${}_2$ and
V10${}_3$, differ at order $1/N_c^2$, so it is not possible to determine which
is more
accurate without computing the $1/N_c^2$ corrections to the isovector magnetic
moments.  The $1/N_c$ relations which are tested by presently available
experimental data are accurate at the level predicted by the $1/N_c$ expansion.
Relations V1, V8${}_1$ and V9${}_1$ which hold to relative order $1/N_c^2$ work
at the $10\%$ level or better.  Relation V10${}_1$ only holds at the $30\%$
level, which is in keeping with a correction of order $1/N_c$.  The $SU(3)$
improved versions of this relation, V10${}_2$ and V10${}_3$, are accurate at
the $10\%$ level or better.  This accuracy is as expected since deviations from
the relations are suppressed either by one power of $1/N_c$ and $SU(3)$
breaking, or by $1/N_c^2$.  The $1/N_c$ expansion predicts that the
experimentally untested relations V2--V7 will hold at the $10\%$ level when the
relevant magnetic moments are measured.

The isovector magnetic moment relations of the $1/N_c$ expansion can be
contrasted with the predictions of the non-relativistic quark model.  Seven of
the relations (V1--V7) are true in the quark model.  There are two additional
relations, V8${}_1$ and V9${}_1$, which are valid to relative order $1/N_c^2$
in the $1/N_c$ expansion.  The quark model has two very similar predictions,
V8${}_2$ and V9${}_2$.  The quark model relation V8${}_2$ works as well its
$1/N_c$ counterpart V8${}_1$.  The quark model relation V9${}_2$, however, is
much less accurate than the $1/N_c$ relation V9${}_1$.\foot{The unmeasured
$n\Delta^0$ transition magnetic moment is eliminated from relation V9 using the
the isospin relation $n\Delta^0=p\Delta^+$.  It is known that the $p\Delta^+$
transition moment is not in good agreement with the quark model prediction
\ref\walker{R.L. Walker, in {\it Proceedings of the 4th International Symposium
on Electron and Photon Interactions at High Energies}, edited by D.~Braben,
(Daresbury Nuclear Physics Laboratory, Daresbury, 1969)\semi L.A.~Copley,
G.~Karl and E.~Obryk, \pl {29} {1969} 117.}.}\ {\it{The failure of the quark
model prediction for the}} $p\Delta^+$ {\it{transition magnetic moment is
resolved by the}} $1/N_c$ {\it{expansion, which gives a different prediction
for the transition magnetic moment.}}  Finally, there is one additional
relation in the quark model, V10${}_4$, which works well, and is the
counterpart of the $1/N_c$ relations V10${}_{1,2,3}$.

The isoscalar magnetic moments (up to corrections of order $1/N_c^2$ ) depend
on the four unknown parameters $a_0$, $a_1$, $b_0$ and $b_1$ in $a(K)$ and
$b(K)$.  Thus, there are six relations among the 10 isoscalar magnetic moments
(S1--S6) which are valid to order $1/N_c^2$ and to all orders in $SU(3)$
breaking. The parameters $a_1$ and $b_1$ arise at first subleading order in the
$1/N_c$ expansion, so there are two additional isoscalar relations (S7 and S8),
which are satisfied by the leading order term to all orders in $SU(3)$ breaking
but which are broken at order $1/N_c$.  Since $b_1=0$ in the $SU(3)$ limit, one
relation (S8) also holds at order $1/N_c$ in the $SU(3)$ limit. The
experimental accuracies of the isoscalar relations agree with the predictions
of the $1/N_c$ expansion. Of the first six relations, only relation S1 is
tested by experimental data.  The accuracy of this relation is consistent with
a correction of order $1/N_c^2$, as predicted by the $1/N_c$ expansion.
Relation S7 which is corrected at order $1/N_c$ (even in the $SU(3)$ limit) is
expected to work at the $30\%$ level; it works to $22\%$.  Relation S8 is
expected to be satisfied at the $10\%$ level since its correction is order
$SU(3)$ breaking/$N_c$.  The relation holds to $7\%$ accuracy.  The $1/N_c$
expansion also predicts that relations S2--S6 will be satisfied at the $10\%$
level when the relevant magnetic moments are measured.

All eight isoscalar relations are true in the quark model.  The quark model,
however, makes no prediction for the relative accuracies of the different
relations.  The $1/N_c$ expansion explains why one of the quark model
predictions works much worse than the others.

The isovector and isoscalar relations V2 and S2 cannot be tested because the
$\Delta^-$ magnetic moment has not been measured. The two relations can be
combined to obtain the prediction for the $\Delta^{++}$ magnetic moment given
in the last line of Table~3. This prediction is valid to two orders in the
$1/N_c$ expansion in the isovector (order $N_c$ and one) and isoscalar (order
one and $1/N_c$) contributions. The experimental measurement of the
$\Delta^{++}$ magnetic moment has a large error so the relation cannot be
tested precisely. However, the present experimental value is consistent with
the theoretical prediction from the $1/N_c$ expansion.

Eighteen relations among the 21 isovector and isoscalar magnetic moments have
been obtained in both the $1/N_c$ expansion and the quark model. In the $SU(3)$
limit, the isoscalar parameter $b_0$ is related to the isovector parameter
$\mu_0$, so there one additional relation (S/V${}_1$) which normalizes the
isoscalar magnetic moments relative to the isovector magnetic moments.
Relation S/V${}_1$ is satisfied by both the order one and $1/N_c$ terms in the
$SU(3)$ limit.\foot{The isovector contribution in S/V${}_1$ has a coefficient
of order $1/N_c$, which makes the order $N_c$ isovector term of the same order
as the order one isoscalar terms.  In deriving S/V${}_1$, we have used the
expression for the total magnetic moment $\mu = -\mu_V + \mu_S/{\sqrt 3}$.  The
minus sign in front of $\mu_3$ is present if one uses the Condon-Shortley phase
convention for the isospin representations.}\ The $1/N_c$ expansion predicts
that this relation is violated by $SU(3)$ breaking at order one in the $1/N_c$
expansion, i.e. at the $\sim 30\%$ level.  The relation works a factor of three
better than this prediction. Although this level of accuracy is not prohibited
by the $1/N_c$ expansion, it would be interesting if this level of accuracy
followed from some other considerations.  We leave this issue as an open
question.  A relation between the isoscalar and isovector magnetic moments can
be obtained in chiral perturbation theory (but not in the $1/N_c$ expansion),
without relying on any models. The relation obtained in ref.~\jlsm, $6 \Lambda
+ \Sigma^- + 4\sqrt3 \Lambda\Sigma^0 = 4 n - \Sigma^+ + 4\Xi^0$ is valid
including all $SU(3)$ breaking corrections of order $m_s^{1/2}$, $m_s \ln m_s$,
and $m_s$, and works to $6 \pm 4\%$.  There is also one additional relation
which relates the isoscalar and isovector magnetic moments in the quark model
if one imposes the isospin constraint $\mu_u=-2\mu_d$.  This relation
(S/V${}_2$): $(p-n)=5(p+n)$, is equivalent to the famous $SU(6)$ prediction
$p/n=-3/2$~\ref\beg{M.A.~Beg, B.W.~Lee, and A.~Pais, \prl{13} {1964} 514.}, and
is satisfied to $7\%$.

In summary, we have derived relations among the baryon magnetic and transition
magnetic moments in the $1/N_c$ expansion.  The $1/N_c$ analysis makes definite
predictions for the accuracies with which these relations are satisfied. With
the notable exception of S/V${}_1$, these predictions are in complete accord
with experiment.  The structure of the $1/N_c$ expansion is much richer than
that of the non-relativistic quark model.  The quark model predicts all the
baryon magnetic moments in terms of three input parameters. Some of the
predictions work better than others. The $1/N_c$ expansion naturally predicts
the hierarchy of relations given in Table~3, and explains which relations work
better than others. There is no particular reason to analyze the magnetic
moments in terms of the relations given in Table~3 in the quark model.

\bigskip
\centerline{{\bf Acknowledgements}}
We would like to thank R.~Dashen for helpful discussions. Recently,
ref.~\ref\lmrw{M.A.~Luty, J.~March-Russell, and M.~White, LBL-35598, {\tt
[hep-ph/9405272]}} has analyzed the baryon magnetic moments in a combined
perturbative expansion in $1/N_c$ and $m_s$ at leading orders. The results
derived in this paper are not obtained in ref.~\lmrw. This work was supported
in part by the Department of Energy under grant number DOE-FG03-90ER40546. A.M.
was also supported by PYI award PHY-8958081.

\listrefs
\listfigs
\insertfig{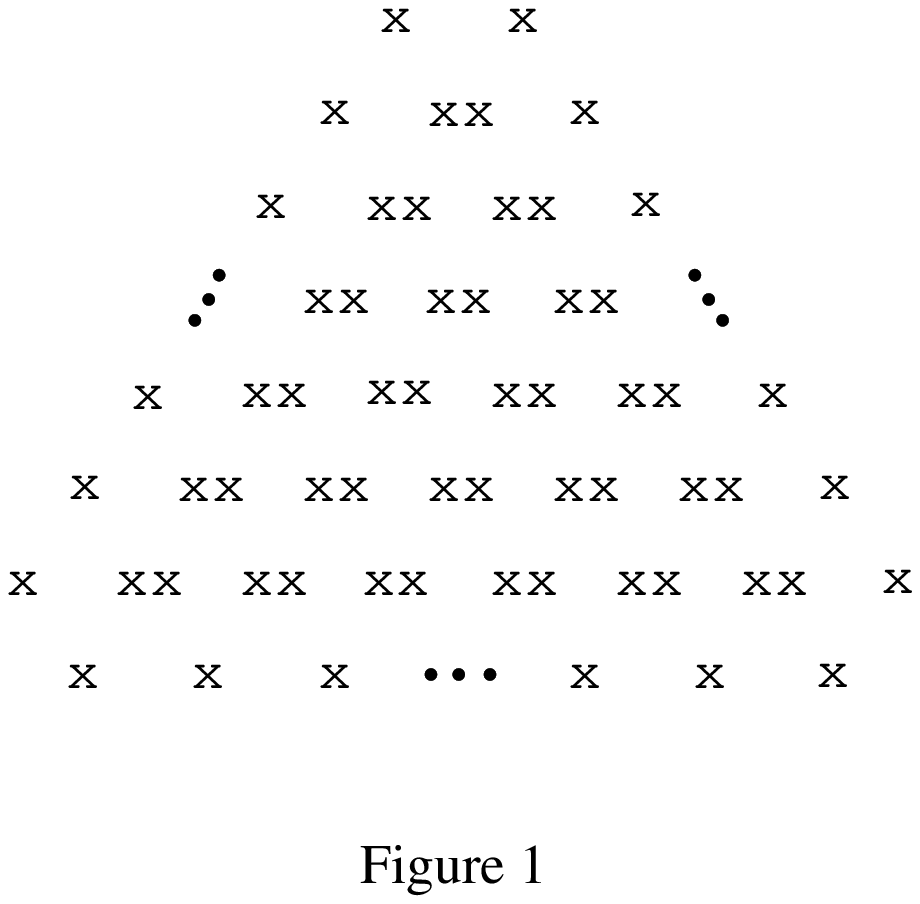}
\vfill\break\eject
\def\frac#1#2{{\textstyle{#1/ #2}}} %puts a small fraction
\def\minus{\phantom{-}}
\centerline{TABLE 1}
\bigskip
\centerline{The matrix elements of the $1/N_c$ and quark model operators for}
\centerline{the baryon magnetic moments.}
\bigskip
\centerline{\vbox{\tabskip=0pt \offinterlineskip
\def\tablerule{\noalign{\hrule}}
\halign{
&\vrule #&\strut\hfil\quad $\rm #$\quad \hfil\cr
\tablerule
&  && a(K) && b(K) && N_c \mu(K) && \mu_u && \mu_d && \mu_s &\cr
\tablerule
& p && \frac12 && \minus 0 && -\frac13&& \minus\frac43 && -\frac13 && \minus 0
&\cr
& n && \frac12 && \minus 0 && \minus \frac13&& -\frac13 && \minus \frac43
&&\minus 0 &\cr
& \Lambda && \frac12 && \minus\frac12 && \minus0 && \minus0 && \minus0 &&
\minus1 &\cr
& \Lambda\Sigma^0 && 0 && \minus0 && \minus\frac13 && -\sqrt\frac13 &&
\minus\sqrt\frac13 && \minus0 &\cr
& \Sigma^+ && \frac12 && -\frac16 && -\frac13&& \minus\frac43 &&\minus 0  &&
-\frac13 &\cr
& \Sigma^0 && \frac12 && -\frac16 && \minus0 &&\minus \frac23 &&\minus \frac23
&& -\frac13 &\cr
& \Sigma^- && \frac12 && -\frac16 &&\minus \frac13&&\minus 0 &&\minus \frac43
&& -\frac13 &\cr
& \Xi^0 && \frac12 &&\minus \frac23 &&\minus \frac19 && -\frac13 &&\minus 0 &&
\minus\frac43 &\cr
& \Xi^- && \frac12 &&\minus \frac23 && -\frac19 &&\minus 0 && -\frac13 &&
\minus\frac43 &\cr
& \Delta^{++} && \frac32 &&\minus 0 && -\frac35 &&\minus 3 &&\minus 0 &&\minus
0 &\cr
& \Delta^{+\minus} && \frac32 &&\minus 0 && -\frac15 &&\minus 2 &&\minus 1
&&\minus 0 &\cr
& \Delta^{0\minus} && \frac32 &&\minus 0 &&\minus \frac15 &&\minus 1 &&\minus 2
&&\minus 0 &\cr
& \Delta^{-\minus} && \frac32 &&\minus 0 &&\minus \frac35 &&\minus 0 &&\minus 3
&&\minus 0 &\cr
& \Sigma^{*+} && \frac32 &&\minus \frac12 && -\frac12 &&\minus 2 &&\minus 0 &&
\minus1 &\cr
&\Sigma^{*0} && \frac32 &&\minus \frac12 &&\minus 0 &&\minus 1 &&\minus 1
&&\minus 1 &\cr
&\Sigma^{*-} && \frac32 &&\minus \frac12 &&\minus \frac12 &&\minus 0 &&\minus 2
&&\minus 1 &\cr
&\Xi^{*0} && \frac32 &&\minus 1 && -\frac13 &&\minus 1 &&\minus 0 &&\minus 2
&\cr
&\Xi^{*-} && \frac32 &&\minus 1 &&\minus \frac13 &&\minus 0 &&\minus 1 &&\minus
2 &\cr
&\Omega^- && \frac32 &&\minus \frac32 &&\minus 0 &&\minus 0 &&\minus 0 &&\minus
3 &\cr
&p\Delta^+&& 0 &&\minus 0 && -\frac{\sqrt2}3 &&\minus \frac{2\sqrt2}3 &&
-\frac{2\sqrt2}3 &&\minus 0 &\cr
&n\Delta^0&& 0 &&\minus 0 && -\frac{\sqrt2}3 && \minus\frac{2\sqrt2}3 &&
-\frac{2\sqrt2}3 &&\minus 0 &\cr
&\Lambda\Sigma^{*0}&& 0 &&\minus 0 && -\frac{\sqrt2}3 &&\minus \sqrt\frac23 &&
-\sqrt\frac23 &&\minus 0 &\cr
&\Sigma\Sigma^{*+}&& 0 && -\frac{\sqrt2}3 && -\frac1{3\sqrt2} &&
\minus\frac{2\sqrt2}3 &&\minus 0 && -\frac{2\sqrt2}3 &\cr
&\Sigma\Sigma^{*0}&& 0 && -\frac{\sqrt2}3 && 0 && \frac{\sqrt2}3 &&
\frac{\sqrt2}3 && -\frac{2\sqrt2}3 &\cr
&\Sigma\Sigma^{*-}&& 0 && -\frac{\sqrt2}3 && \frac1{3\sqrt2} && 0 &&
\frac{2\sqrt2}3 && -\frac{2\sqrt2}3 &\cr
&\Xi\Xi^{*0}&& 0 && -\frac{\sqrt2}3 && -\frac{2\sqrt2}9 && \frac{2\sqrt2}3 && 0
&&-\frac{2\sqrt2}3 &\cr
&\Xi\Xi^{*-}&& 0 && -\frac{\sqrt2}3 && \frac{2\sqrt2}9 && 0 && \frac{2\sqrt2}3
&& -\frac{2\sqrt2}3 &\cr
\tablerule
}}}

\vfill\break\eject

\def\frac#1#2{{\textstyle{#1\over #2}}} %puts a small fraction
\centerline{TABLE 2}
\bigskip
\centerline{Isospin relations among the baryon magnetic moments.}
\bigskip
\centerline{\vbox{ \tabskip=0pt \offinterlineskip
\def\tablerule{\noalign{\hrule}}
\def\space{height 2pt&\omit&&\omit&&\omit\cr}
\halign{
\vrule #&\strut\hfil\ #\hfil&&
\vrule #&\strut\hfil\ $\rm #$\ \hfil\cr
\tablerule
&    && \omit \hfil Isospin Relations\hfil&&\omit\cr
\tablerule\space
& I1 && \Sigma^{+} + \Sigma^-= 2 \Sigma^0
&&\omit\cr
& I2 && \Sigma^{*+}+\Sigma^{*-} = 2
\Sigma^{*0}&&\omit\cr
& I3 && \Sigma\Sigma^{*+}+\Sigma\Sigma^{*-} =
2 \Sigma\Sigma^{*0} &&\omit\cr
& I4 &&
\Delta^{++}-\Delta^-=3(\Delta^+-\Delta^0)&&\omit\cr
& I5 &&
\Delta^{++}+\Delta^-=\Delta^++\Delta^0&&\omit\cr
& I6 && p\Delta^{+}= n\Delta^0&&\omit\cr
\space\tablerule
}}}
\vskip2in
\centerline{TABLE 3}
\bigskip
\narrower
Relations among the baryon magnetic moments in the $1/N_c$ expansion, and in
the non-relativistic quark model. The isovector magnetic moments are of order
$N_c$, and the isoscalar magnetic moments are of order one. A $\surd$ implies
that the relation is satisfied to that order in $1/N_c$ {\it to all orders in
$SU(3)$ breaking}. A $SU(3)$ implies that the relation is satisfied to that
order in $1/N_c$ only in the $SU(3)$ limit. The experimental accuracies are
given in the last column for the relations whose magnetic moments have been
measured.
\vfill\break\eject
\centerline{\vbox{ \tabskip=0pt \offinterlineskip
\def\tablerule{\noalign{\hrule}}
\def\space{height 2pt&\omit&&\omit&&\omit&&\omit&&\omit&&\omit&&\omit\cr}
\halign{
\vrule #&\strut\hfil\ #\hfil&&
\vrule #&\strut\hfil\ $\rm #$\ \hfil\cr
\tablerule\space\tablerule
&    && \omit \hfil Isovector Relations\hfil && N_c   && 1 && {\rm QM}
&&\omit&&\omit\cr
\tablerule\space\tablerule\space
& V1 && (p-n)-3(\Xi^0-\Xi^-)=2(\Sigma^{+} -
\Sigma^-)&&\surd&&\surd&&\surd&&10\pm2\%&&\omit\cr
& V2 && \Delta^{++}-\Delta^-=\frac95(p-n)
&&\surd&&\surd&&\surd&&\omit&&\omit\cr
& V3 &&
\Lambda\Sigma^{*0}=- \sqrt2 \Lambda \Sigma^{0}&& \surd&& \surd&& \surd&&
\omit&& \omit\cr
& V4 && \Sigma^{*+}-\Sigma^{*-} = \frac32 (\Sigma^{+} - \Sigma^-)
&&\surd&&\surd&&\surd&&\omit&&\omit\cr
& V5 && \Xi^{*0}-\Xi^{*-} = -3(\Xi^0-\Xi^-)
&&\surd&&\surd&&\surd&&\omit&&\omit\cr
& V6 && \sqrt2(\Sigma\Sigma^{*+}-\Sigma\Sigma^{*-}) =
(\Sigma^{+} - \Sigma^-) &&\surd&&\surd&&\surd&&\omit&&\omit\cr
& V7 && \Xi\Xi^{*0}-\Xi\Xi^{*-}=-2\sqrt2(\Xi^0-\Xi^-)
&&\surd&&\surd&&\surd&&\omit&&\omit\cr
\tablerule
& V8${}_1$ && -2\Lambda\Sigma^0=(\Sigma^+-\Sigma^-)&&\surd&&\surd&& No
&&11\pm5\%&&\omit\cr
& V8${}_2$ && -2\Lambda\Sigma^0=\frac{\sqrt3}{2}(\Sigma^+-\Sigma^-)&&No&&No&&
\surd&&4\pm5\%&&\omit\cr
& V9${}_1$ && p\Delta^++n\Delta^0=\sqrt2(p-n)&&\surd&&\surd&& No
&&3\pm3\%&&\omit\cr
& V9${}_2$ && p\Delta^++n\Delta^0=\frac{4\sqrt2}{5}
(p-n)&&No&&No&&\surd&&26\pm4\%&&\omit\cr
\tablerule
& V10${}_1$ &&(\Sigma^+-\Sigma^-)=(p-n)&&\surd&&No&&No&&27\pm1\%&&\omit\cr
& V10${}_2$ &&(\Sigma^+-\Sigma^-)=\left(1-{1\over N_c}\right)(p-n)
&&\surd&&SU(3)&&No&&13\pm2\%&&\omit\cr
& V10${}_3$ &&\left(1+{1\over
N_c}\right)(\Sigma^+-\Sigma^-)=(p-n)&&\surd&&SU(3)&&No&&1\pm2\%&&\omit\cr
& V10${}_4$
&&(\Sigma^+-\Sigma^-)=\frac45(p-n)&&No&&No&&\surd&&5\pm2\%&&\omit\cr
\space\tablerule\space\tablerule
&    && \omit \hfil Isoscalar Relations\hfil && 1   && 1/N_c && {\rm QM}
&&\omit&&\omit\cr
\tablerule\space\tablerule\space
& S1 && (p+n)-3(\Xi^0+\Xi^-) = - 3 \Lambda + \frac32(\Sigma^{+} +
\Sigma^-)-\frac43\Omega^-&&\surd&&\surd&&\surd&&4\pm5\%&&\omit\cr
& S2 && \Delta^{++}+\Delta^-=3(p+n) &&\surd&&\surd&&\surd&&\omit&&\omit\cr
& S3 && \frac23 (\Xi^{*0}+\Xi^{*-})=\Lambda+\frac32(\Sigma^{+} + \Sigma^-)
-(p+n)+(\Xi^0+\Xi^-)&&\surd&&\surd&&\surd&&\omit&&\omit\cr
& S4 && \Sigma^{*+}+ \Sigma^{*-} =
\frac32(\Sigma^{+} + \Sigma^-)+3\Lambda &&\surd&&\surd&&\surd&&\omit&&\omit\cr
& S5 && \frac{3}{\sqrt2}(\Sigma\Sigma^{*+}+\Sigma\Sigma^{*-}) =
3(\Sigma^{+} +\Sigma^-)- (\Sigma^{*+} +\Sigma^{*-})
&&\surd&&\surd&&\surd&&\omit&&\omit\cr
& S6 && \frac{3}{\sqrt2}(\Xi\Xi^{*0}+\Xi\Xi^{*-})=-3(\Xi^0+\Xi^-)
+(\Xi^{*0}+\Xi^{*-})&&\surd&&\surd&&\surd&&\omit&&\omit\cr
\tablerule
&S7&&5(p+n)-(\Xi^0+\Xi^-)=4(\Sigma^{+}+\Sigma^-)
&&\surd&&No&&\surd&&22\pm4\%&&\omit\cr
&S8&&(p+n)-3\Lambda=\frac12(\Sigma^{+}+\Sigma^-)-(\Xi^0+\Xi^-)
&&\surd&&SU(3)&&\surd&&7\pm1\%&&\omit\cr
\space\tablerule\space\tablerule
&    && \omit \hfil Isoscalar/Isovector Relations\hfil && 1   && 1/N_c && {\rm
QM} &&\omit&&\omit\cr
\tablerule\space\tablerule\space
&S/V${}_1$ && (\Sigma^++\Sigma^-)-\frac12(\Xi^0+\Xi^-) = \frac12(p+n)+
3\left({1\over N_c}-{2\over
N_c^2}\right)(p-n)&&SU(3)&&SU(3)&&No&&10\pm3\%&&\omit\cr
&S/V${}_2$ && p-n = 5(p+n)&&No&&No&&\surd&&7\%&&\omit\cr
\space\tablerule\space\tablerule\space
&
&&\Delta^{++}=\frac32(p+n)+\frac{9}{10}(p-n)&&\surd&&\surd&&\surd&&
21\pm10\%&& \omit\cr
\space\tablerule\space\tablerule
}}}

\bye